# The Detection and Localization of Frost Protein in *Drosophila*

**Haroun Mansuar**

The University of Guelph

Department of Molecular and Cellular Biology

**Abstract**

In response to cold stress, *Drosophila Melanogaster* increase their expression of Frost, a candidate gene involved in cold response. The direct role of Frost in cold tolerance is yet to be determined, and its importance for survival in cold environments has been questioned. In this study, I attempt to better understand the molecular machinery of Frost by selecting for its protein in fly lysate knowing only its RNA has been found. Detection of Frost expression and subsequently studying its protein will hopefully lead to enhanced comprehension of cold responses in insects, which is the long-term goal of this research. I predict that Frost will be expressed in flies that undergo cold stress at 0°C for 2 hours before recovering at $22^0$C for 3 hours before lysing. Two Western blots were executed using fly lysate that underwent the above treatment, whose antibodies were specific to Frost. No bands were seen in any of the lanes containing treated samples in either of the Immunoblots, indicating that Frost was either not expressed or not detected in fly lysate.

**Introduction**

The distribution and abundance of insects are largely affected by temperature. As temperatures plummet, insects undergo a reversible process called chill coma, where they temporarily lose the ability to move (Denlinger and Lee, 2010). As such, Chill susceptibility is a term used for species that succumb to cold exposure without the formation of ice crystals (Denlinger & Lee, 2010). However, it was found that for many insect species, exposure to a nonlethal short lived cold stress was beneficial, as it raised tolerance levels for more extreme cold stresses (Denlinger & Lee, 2010). This was termed rapid cold hardening and is essential to the survival of insect species in colder environments (Denlinger and Lee, 2010). Due to its varying cold tolerances and phenotypic plasticity, *Drosophila Melanogaster* is used as the genetic model species for studies involving the effects of cold exposures (Udaka *et al*. 2013).

Genes that show increased up-regulation after cold exposure may contribute in the enhancement of cold tolerance. Previous studies have found a gene whose expression was considerably increased after exposure to cold in *Drosophila melanogaster,* which they called Frost (Goto, 2001). Consisting of 278 amino acids, Frost appears to be a stress-related disordered protein whose modular structure indicates that it is directed into the endoplasmic reticulum and secreted into the extracellular space (Goto, 2001). Other notable characteristics include a proline rich region and an unusually high number of internal repeats in both amino acid and nucleotide sequences (Goto, 2001).

The exact mechanism by which Frost enhances cold response remains unknown and its importance for cold tolerance is being questioned. Conflicting reports have emerged concerning whether *Drosophila* needs to express frost to survive cold environments. In one study, RNAi was used to knock down Frost expression in fruit flies which were subsequently exposed to cold but yielded a similar number of survivors regardless of Frost expression levels (Udaka *et al*. 2013). Another study, testing the same hypotheses using a similar methodology found that the absence of Frost was lethal in *Drosophila* when exposed to colder conditions (Colinet *et al*. 2010). The discrepancy in the results led researchers to question whether different life stages in flies had varying cold tolerances. The answer was a resounding yes, as studies have shown that significant up-regulation of Frost occurs in eggs, third instar larvae, 2 and 5 day old flies (Bing *et al*. 2012). Also, much like in heat tolerance, cold tolerance sees a rapid decline as flies age indicating that studies should be conducted on flies in the same life stage (Colinet *et al*. 2013). While stress genes Hsp22 and Hsp70Aa saw a similar decline in function with increasing age, previous studies have concluded that no correlation is evident between stress genes and cold tolerance (Colinet *et al*. 2013; Udaka *et al.* 2010).

Clearly, consensus on the importance and function of Frost is difficult to come by, as the genetic machinery by which it works remains, for the most part, a mystery. Additionally, only its RNA has been identified, making the task of piecing together this protein's function more challenging. This study will attempt to select for Frost protein in fly lysate. We predict that, under optimized conditions that includes both cold shock and recovery, Frost protein will be expressed in fly lysate if it is essential for cold tolerance and will therefore be detected by an immunoblot.

**Results**

In order to verify the expression of Frost in fly lysate, a Western Blot was run whose antibodies were specific to the protein of interest. Flies that did not undergo any special treatment were used as a cold treatment control and were located next to a lane containing just recombinant Frost, another control testing for the antibody specificity to Frost. *Drosophila Melanogaster* that underwent cold treatment for 2 hours at $0^{\circ}\text{C}$ and allowed to recover for 3 hours at room temperature was the treated sample that was predicted to yield the frost protein. Using BRMW protein ladder, the aforementioned samples were underwent the aforementioned Western Blot (Figure 2). Frost was detected in the recombinant Frost lane, as expected, but not in the treated samples regardless of the concentration of fly lysate used (Figure 2). A band can be detected in the control lane containing untreated fly lysate (Figure 2).

A subsequent Immunoblot was run with the same samples in the same order as done previously. After samples were separated by molecular weight through SDS-PAGE, the gel was stained with

Commassie-Blue to ensure that the samples were present and separated effectively (Figure 1). As can be seen, the ladder, control and treated samples were present and seemed to have ran properly (Figure 1). The lane containing Frost was not present, as the amount of the protein used was too small to be detected by an SDS-PAGE (Figure 1).

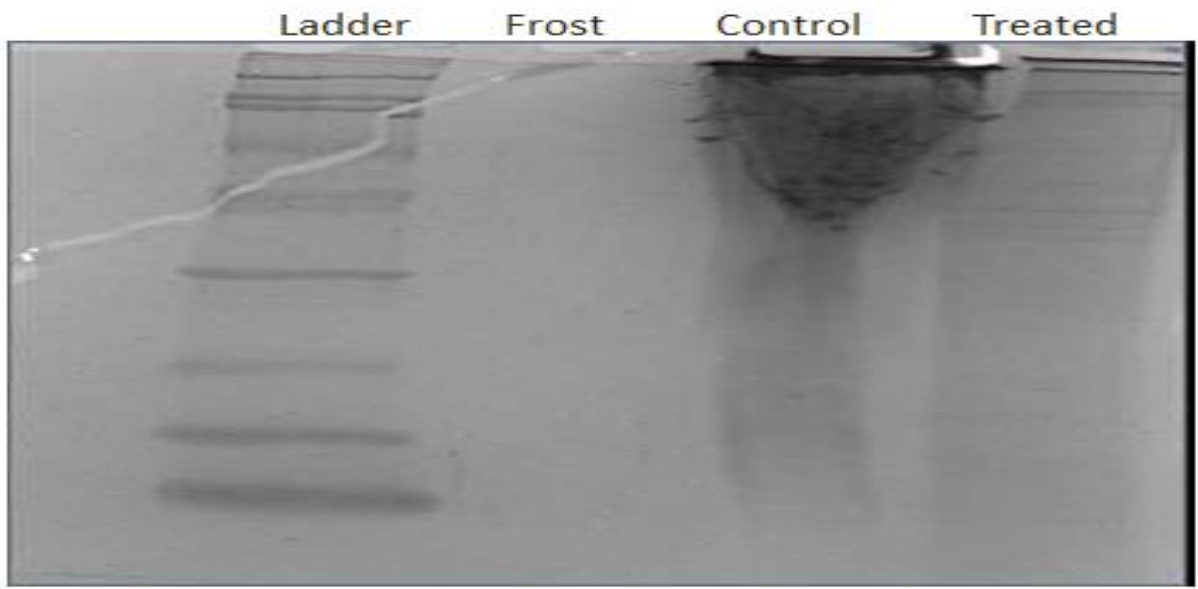

**Figure 1.** A stain of SDS-PAGE containing lanes with BRMW ladder, recombinant Frost, untreated and treated fly lysate. The samples were run at 100 V for 45 minutes then 125V for an hour.

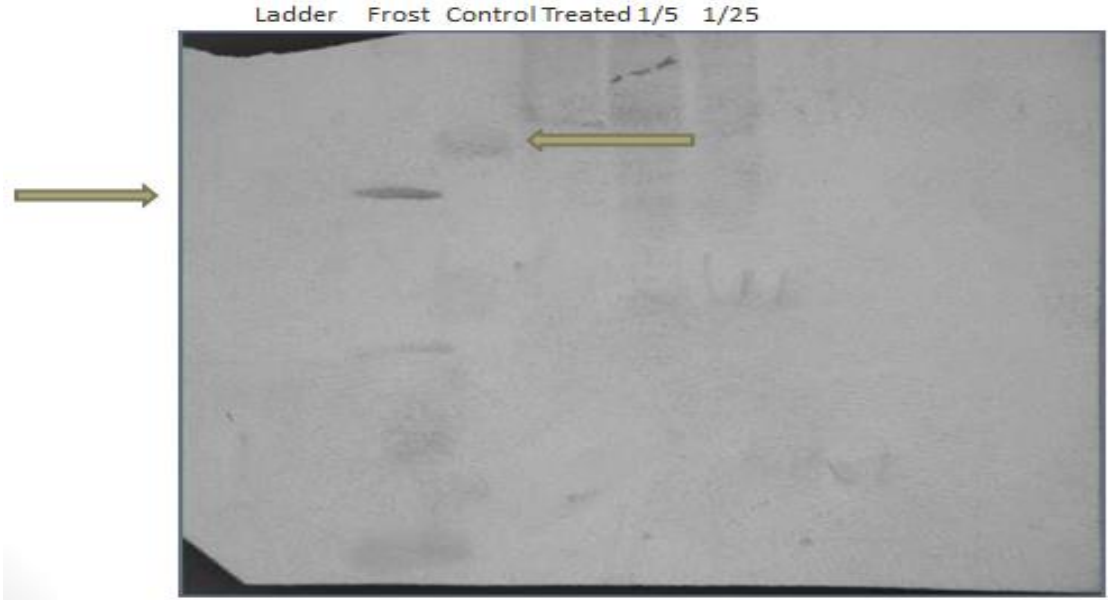

**Figure 2.** Western Blot using BRMW ladder, recombinant Frost and untreated fly lysate as a control and three treated lysates with different concentrations (1:1, 1:5, 1:25). As per usual Western Blot protocol outlined in the Methods section, samples were run on SDS-PAGE for 45 minutes at 100V and 1 hr at 125V. Proteins were then transferred for 1 hr at 100v and 350mA before being probed with antibodies and imaged.

Due to the fact that the samples were observed when stained, it was deemed logical to continue with the Immunoblot procedure. The imaging of the membrane yielded no bands in either the treated or untreated lysate samples (Figure 3). There was, however, a band in the recombinant Frost lane, indicating that positive control worked (Figure 3).

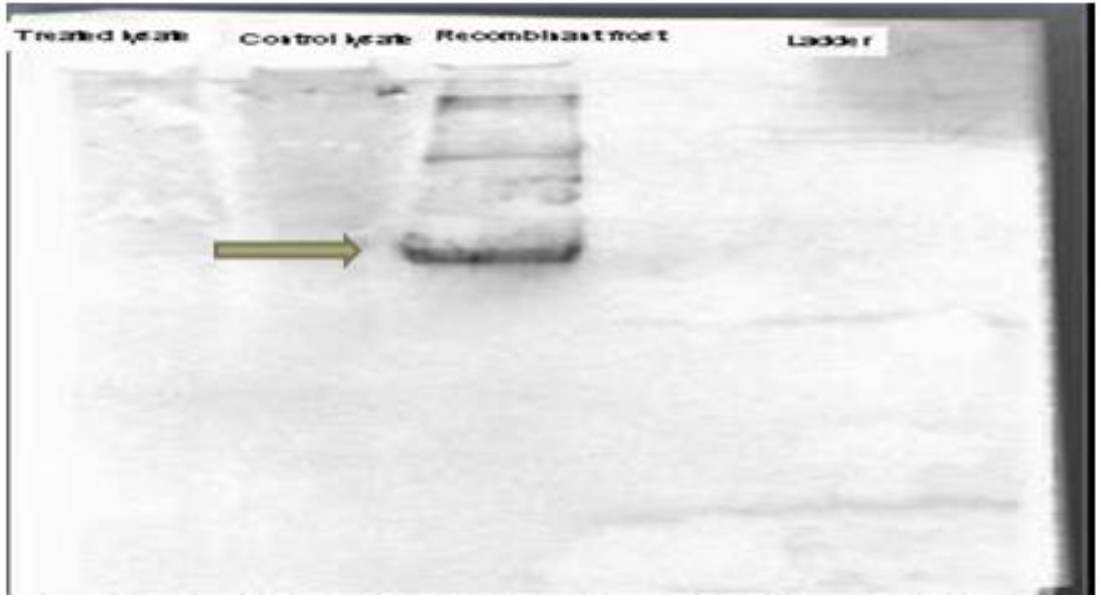

**Figure 3.** Imaged membrane for previous gel. From left to right, the lanes contained treated and untreated fly lysate, recombinant frost, and BRMW ladder. After SDS-PAGE, the samples were transferred to a membrane at 100V, 350 mA for an hour before being probed by a primary, then secondary antibody before chromogenic imaging.

**Discussion**

In conclusion, while the results presented in this research are preliminary, the fact that the Frost protein was not detected in samples that underwent cold treatment and recovery disproves our hypotheses. Theoretically, the treatment that the *Drosophila Melanogaster* were exposed to (2 hours at 0 degrees before recovery) were similar to the conditions that were presented in the literature, which then should have yielded the protein of interest (Udaka *et al.* 2013). However, two attempts were made to select for Frost, neither of which was successful as no bands were seen in the lanes containing the treated fly lysate (Figure 1; Figure 3). The fact that Frost protein was detected in control lanes containing recombinant Frost shows that the protocol for the immunoblot is effective and does not need further optimized (Figure 1; Figure 3). Interestingly, a band was detected in the lane containing untreated fly lysate (Figure 1). This was surprising as this was a negative control lane in which we did not expect to see any Frost present. While the band was not the same size as the Frost protein on the membrane, it is possible that N-terminal glycosylation of the protein may have slowed the protein and did not allow it to travel down the membrane as far as Frost usually does (Figure 1). In terms of the absence of Frost in the treated Fly lysate, much remains to be answered. While the treatments for the flies are known, the age at which the flies were lysed and experimented on is not. As the literature suggested, there are life stages in *Drosophila Melanogaster* that express Frost more than others (Bing *et al.* 2012). Still, the fact remains that only the RNA of Frost has been found, while its protein has not despite treatment conditions that should warrant its expression.

Future studies involving the detection of Frost protein should start with the identification of the band that was seen in the untreated fly lysate (Figure 1). This band could be Frost, an artifact, or something completely unbeknownst to us. Verifying the specificity of the antibody we used should be the first step to determine if what was seen on the membrane really the protein of interest. Elution of the protein should follow, and will be run again on the SDS-PAGE to compare its size with Frost. Finally, sequencing of the protein to determine its identity would be another option.

**Materials and Methods**

*Insect sample preparation*

80 mg (equivalent to 100 flies) of *Drosophila Melanogaster* underwent cold treatment at $0^0$C for 2 hours before recovering at $22^oC$ for 3 hours. These flies were then lysed and stored at $-80^0$C until usage. Another 80mg of flies, which underwent no treatment, were lysed and stored at the same temperature. Both samples were lysed by bead beating for 20 seconds at max speed by 250uL Zirconial Silica Beads before placing on ice for 30 seconds. This was repeated 5 times. 500uL of 50mM of sodium phosphate buffer ( detergent, pulls things off membrane) was then added. Samples were then centrifuged down at 14.5rpm for 15 minutes before supernatant was collected.

*Western Blotting*

In order to separate the proteins based on molecular weight, the protein samples, which consist of BRMW ladder, recombinant Frost, and the treated and control fly lysate (as described above) were run on an SDS-PAGE. Samples were run at 100V for 45 minutes before the voltage is increased to 125V for an hour. Upon completion, the gel was stained using coomassie blue R-250 to ensure samples were present and ran properly. The gel was washed several times before staining occurred for 5 minutes and de-staining followed for 10 minutes.

The separated proteins are then transferred to a PVDF membrane by first assembling the "transfer sandwich" where the protein transfer occurred. Careful attention was paid to the orientation of the transfer sandwich. The sandwich is then placed into the same apparatus that the samples were first separated, along with pre-made transfer buffer, an ice pack and a small magnetic spinner. The transfer occurred at 100V and 350mA for an hour. After the transfer was completed, the membrane was then submerged in 3% blocking solution and left overnight. The next day, the membrane was probed with the primary antibody specific to Frost (1uL PAC in 5ml BSA) for 1 hour. The membrane was then washed for 10 minutes with shaking by 5mL TBSTT twice. The membrane was washed again with 5mL TBS for 10 minutes. The secondary antibody (anti-rabbit) was then used to probe for the primary antibody (0.667uL antibody to 5mL BSA) and left on a shaker for an hour. The washing steps described above were repeated. Finally, to image the membrane, Chromogenic AP-based substrate was used as 17uL of BCIP, 33uL NBT and 5mL of AP-substrate was added to the membrane and shaken till bands appeared before a dark purple background.

**Acknowledgements**

I would like to thank Dr. Steffen Graether for agreeing to supervise my research and being an excellent and patient mentor throughout the duration of this project. I would also like to thank my fellow lab mates from both the Graether and Kimber lab, Matthew Clarke in particular, for their encouragement and assistance as this project would not be possible without them.